\documentclass[final,nomathfonts]{aipproc}

\layoutstyle{6x9}

\usepackage{amsfonts,amssymb,color}
  \usepackage[
    colorlinks,%
    linkcolor=blue,citecolor=red,urlcolor=blue,
    hyperindex,%
    plainpages=false,%
    bookmarksopen,%
    bookmarksnumbered%
  ]{hyperref}
\usepackage[all,knot]{xy}
\usepackage{tikz}    

\xyoption{ps}

\newcommand\beq{\begin{equation}}
\newcommand\eeq{\end{equation}}
\newcommand\be{\begin{equation}}
\newcommand\ee{\end{equation}}
\newcommand\beqa{\begin{eqnarray}}
\newcommand\eeqa{\end{eqnarray}}
\newcommand\bean{\begin{eqnarray*}}
\newcommand\eean{\end{eqnarray*}}

\newcommand{\cO}{{\mathcal O}}

\newcommand{\cS}{{\mathcal S}}
\newcommand{\cT}{{\mathcal T}}

\newcommand\U{{\mathrm U}}
\newcommand\C{{\mathbb C}}

\newcommand\R{{\mathbb R}}
\newcommand\Z{{\mathbb Z}}

\newcommand\G{{\mathcal{G}}} 
\newcommand\E{{\mathcal{E}}} 

\newcommand{\SO}{{\rm SO}}

\newcommand{\Hhat}{\hat H}

\newcommand\Irrep{{\rm{Irrep}}}

\newcommand\tensor{\otimes}
\newcommand\To{\Rightarrow}

\newcommand\iso{\cong}

\newcommand\extd{\mathrm {d}}

\newcounter{letter} \newcounter{numeral} \newcounter{Numeral}

\newcommand\maps{\colon}

\newenvironment{proof.within.proof}
{\noindent{\it Proof:}}{
\hfill $\Box$ \medskip}

\newcommand{\rtriangle}{ \begin{tikzpicture}[scale=1.2]
\draw (0,0) -- ++(0,0.16) -- ++(0.1,-0.08) --cycle ;
\end{tikzpicture}}

\newcommand{\utriangle}{\begin{tikzpicture}[scale=1.2]
\draw (0,0) -- ++(0.1,0.08) -- ++(0.1,-0.08) --cycle ;
\end{tikzpicture}}

\newcommand{\dtriangle}{\begin{tikzpicture}[scale=1.2]
\draw (0,0) -- ++(0.1,-0.075) -- ++(0.1,0.075) --cycle ;
\end{tikzpicture}}

\newcommand{\ltriangle}{ \begin{tikzpicture}[scale=1.2]
\draw (0,0) -- ++(0,0.16) -- ++(-0.1,-0.08) --cycle ;
\end{tikzpicture}}

\newcommand{\hdiamond}{\begin{tikzpicture}[scale=1.2]
\draw (0,0) -- ++(0.1,0.08) -- ++(0.1,-0.08) --cycle ;
\draw (0,0) -- ++(0.1,-0.08) -- ++(0.1,0.08)  ;
\end{tikzpicture}}

\newcommand{\hvdiamond}{\begin{tikzpicture}[scale=1.4]
\draw (0,0) -- ++(0,0.16) -- ++(0.1,-0.08) -- ++(-0.2,0)--cycle ;
\draw (0,0) -- ++(0.1,0.08) ;
\draw (-0.1,0.08) -- ++(0.1,0.08) ;
\end{tikzpicture}}

\begin{document}

\title{2-Group Representations for Spin Foams}


\author{Aristide Baratin}{
  address={Max Planck Institute for Gravitational Physics, Albert Einstein Institute, Am M\"uhlenberg 1, 14467 Golm, Germany},   
  email={Aristide.Baratin@aei.mpg.de}
}

\author{Derek K.\ Wise}{
  address={Department of Mathematics, University of California, Davis, 95616, USA},
  email={derek@math.ucdavis.edu}
}

\begin{abstract}
Just as 3d state sum models, including 3d quantum gravity, can be built using categories of group representations, `2-categories of 2-group representations' may provide interesting state sum models for 4d quantum topology, if not quantum gravity.  Here we focus on the `Euclidean 2-group', built from the rotation group $\SO(4)$ and its action on the translation group $\R^4$ of Euclidean space. We explain its infinite-dimensional unitary representations, and construct a model based on the resulting representation 2-category. This model, with clear geometric content and explicit `metric data' on triangulation edges, shows up naturally in an attempt to write the amplitudes of ordinary quantum field theory in a background independent way.
\end{abstract}

\maketitle


\section{Introduction}

The success of combinatorial and algebraic methods in 3d quantum gravity \cite{PonzanoRegge,TuraevViro} has long been an inspiration for analogous 4d models, including spin foam models of quantum gravity.  A mathematically elegant approach to getting 4d models from 3d ones uses so called `higher-dimensional algebra'.  Our aim here is not only to explain an instance of this approach, but also to present evidence that the resulting models may be relevant for real-world physics.  Indeed, as we shall explain, they have already shown up in an unexpected way, in an attempt to understand a certain `limit' of quantum gravity.

The reason for the term `higher-dimensional algebra' is easily explained using the example most relevant to this paper: `2-groups' \cite{BaezLauda-2gps}.   Whereas a group might consist of symmetry transformations of some `object' $\star$, drawn as `arrows':
\[
\quad \xymatrix{\star \ar[r]^{g} & \star }
\]
a `2-group' also has {\em `symmetries between symmetries'}:  
\beq
\label{2-arrow}
  \xymatrix{
  \star\ar@/^2ex/[rr]^{g}="g1"\ar@/_2ex/[rr]_{g'}="g2"&&\star
  \ar@{=>}^{h} "g1"+<0ex,-2ex>;"g2"+<0ex,2ex>
}
\eeq 
drawn as `2-arrows'.  This added structure gives 2-groups an additional `algebraic dimension' ordinary groups do not have.  The 2-arrows have two distinct notions of `product', as explained later, and these must satisfy certain `coherence laws' governing their algebraic structure.  Like other instances of higher-dimensional algebra, passing from groups to 2-groups is an example of `categorification', where we have replaced the {\em set} of arrows with the `{\em category}' of arrows and 2-arrows. 

Going up in dimension often goes hand-in-hand with categorification.  
For example, in constructing topological invariants that are calculated using triangulations, a key step is showing invariance under the local `Pachner moves' that allow one to pass between any two triangulations of the same manifold.   Pachner moves are subtly but strongly tied to the coherence laws of higher categories, in such a way that appropriate labels from higher-dimensional algebra can give manifest Pachner invariance.  A recent review by Baez and Lauda \cite{BaezLauda} nicely explains the relationship between physical and `algebraic' dimensions in topological field theory, and the history of this idea, with detailed references. 

Of course, quantum gravity, having local degrees of freedom, may be more than mere quantum topology, but it is certainly not {\em less} than quantum topology.  While `categorifying' 3d models to get 4d ones is unlikely to miraculously yield quantum gravity, the evidence from topology suggests climbing up from 3 to 4 dimensions may require rethinking what sort of mathematics is needed. We should at least consider the possibility that standard quantum gravity approaches will fail as long as they are attempting to solve a {\em four}-dimensional problem using `{\em three}-dimensional' mathematics.

In this note, following Barrett and Mackaay \cite{BarrettMackaay}, as well as Crane, Sheppeard and Yetter \cite{CraneSheppeard,CraneYetter}, we propose using 2-categories of 2-group representations to construct state sum models in dimension 4.  We give an explicit construction using a categorical analog of the Poincar\'e group.  Most importantly, this state sum corresponds precisely to the background independent formulation of ordinary quantum field theory amplitudes derived in \cite{BaratinFreidel1b}.  Our brief treatment here is an exposition of results from \cite{2Rep, BaratinFreidel1b} and the forthcoming papers \cite{BaratinFreidel2, BaratinWise}, to which we refer the reader for further details.

\section{2-Group representations for state sums}

Our goal here is to use the representation theory of 2-groups to construct four dimensional state sum models.  In these models, edges in a triangulation are labeled by representations, and triangular faces are labeled by intertwiners relating the representations on their bounding edges. But 2-group representation theory also involves a notion of `2-intertwiner' between intertwiners, and these 2-intertwiners label tetrahedra.  The 2-category of representations, intertwiners, and 2-intertwiners has been constructed explicitly in \cite{2Rep}.  Here, we only explain enough of the resulting geometric structure to understand the proposed state sum models. 

To describe 2-group representations, we first need a precise algebraic characterization of 2-groups themselves.  Actually, we shall not need the most general sort of 2-groups;  what we shall use are called `strict skeletal 2-groups', which we henceforth simply call {\bf 2-groups}, without qualification.
This is a significantly restricted class of 2-groups, but it includes our main example, and has the advantage that any 2-group of this sort can be constructed from two simple and familiar pieces of data:  
\begin{itemize}
\setlength{\itemsep}{0em}
\item a group $G$  
\item an abelian group $H$ equipped with an action of $G$ as automorphisms.
\end{itemize}
We write elements of $G$ as $g,g',\ldots$, elements of $H$ as $h,h',\ldots$, and the action of $g$ on $h$ as $gh$. 
From these data, we build a 2-group with arrows $\star \to \star$ labeled by $g$'s, and for each $g$, 2-arrows $g\To g$ labeled by $h$'s.  The 2-arrows, usually drawn as in (\ref{2-arrow}), have two algebraic operations, $\cdot$ and $\circ$, called {\bf vertical} and {\bf horizontal multiplication}, for diagrammatically obvious reasons:  
\[
\begin{array}{cccccc}
& & & & & \\[-2em]
(g',h')\cdot(g,h) =&
\xymatrix{
   \star\ar@/^3.5ex/[rr]^{g}="g1"\ar[rr]|-(0.25){g}\ar@{}[rr]|{}="g2"
  \ar@/_3.5ex/[rr]_{g}="g3"&&\star
  \ar@{=>}^{h} "g1"+<0ex,-2ex>;"g2"+<0ex,1ex>
  \ar@{=>}^{h'} "g2"+<0ex,-1ex>;"g3"+<0ex,2ex>
}
&=&
\xymatrix{
 \star\ar@/^3ex/[rr]^{g}="g1"
  \ar@/_3ex/[rr]_{g}="g3"&&\star
  \ar@{=>}^{h'h} "g1"+<0ex,-2ex>;"g3"+<0ex,2ex>
}
& = (g,h'h)
 \\[.5em]
 (g',h')\circ(g,h) =&
\xymatrix@C=1.5em{
  \star\ar@/^2ex/[rr]^{g}="g1"\ar@/_2ex/[rr]_{g}="g2"&&\star
  \ar@{=>}^{h} "g1"+<0ex,-2ex>;"g2"+<0ex,2ex>
  \ar@/^2ex/[rr]^{g'}="g1"\ar@/_2ex/[rr]_{g'}="g2"&&\star
  \ar@{=>}^{h'} "g1"+<0ex,-2ex>;"g2"+<0ex,2ex>
}
&=&
\xymatrix@C=2.5em{
 \star\ar@/^2ex/[rr]^{g' g}="g1"
  \ar@/_2ex/[rr]_{g' g}="g3"&&\star
  \ar@{=>}^{h'(g' h)} "g1"+<-2ex,-2ex>;"g3"+<-2ex,2ex>
}
& = (g'g,h'(g'h))
\end{array}
\]
We note that from the data $(G,H)$, we could instead simply construct the semidirect product $G\ltimes H$.  While this group is  involved in the horizontal multiplication, 2-groups have a richer algebraic structure, as well as a richer representation theory.

The {\bf Poincar\'e 2-group} is the 2-group for which $G$ is the Lorentz group, and $H$ is the group of translations of Minkowski space \cite{Baez1}.  For simplicity, we work instead with its positive-signature analog, the {\bf Euclidean 2-group} $\E$, with $G=\SO(4)$, $H=\R^4$.  We now describe both the representation theory of 2-groups \cite{2Rep}, using the Euclidean 2-group as an example, and how the representation theory can be used in state sum models. 

{\bf Edges: representations.}  Edges in our state sum models are labeled by `irreducible representations'.  Any (unitary, measurable) representation of $\E$ is given by an  $\SO(4)$-equivariant map $\chi\maps X \to \R^4$, where $X$ is some space on which the rotation group acts. Representations for which the action on $X$ is transitive are {\bf indecomposable} representations. {\bf Irreducible} representations are indecomposable ones for which the map $\chi$ is one-to-one, in which case $X$ is isomorphic to a single $\SO(4)$-orbit in $\R^4$, a 3-sphere of given radius. So, edges are effectively labeled by positive numbers, the radii of spheres.

{\bf Gluing edges: tensor products.}  Joining two labeled edges at an endpoint: 
\(
\xy
     (-3,-1)*{\scriptscriptstyle\bullet}="A";
     (0,1)*{\scriptscriptstyle\bullet}="B";
     (3,-1)*{\scriptscriptstyle\bullet}="C";
     "A";"B"**\dir{-};
     "C";"B"**\dir{-};
     (-2,1.5)*{\scriptscriptstyle X\phantom{'}};
     (3,1.5)*{\scriptscriptstyle X'};
\endxy
\) 
means taking the `tensor product' of representations.  The {\bf tensor product} $X\tensor X'$ of two irreps of $\E$ turns out to correspond to the map $X \times X'\to \R^4$ given by $(x,x')\mapsto x+ x'$.   

{\bf Triangles: intertwiners.} Triangular faces: \(
\xy
     (-3,-1)*{\scriptscriptstyle\bullet}="A";
     (0,1)*{\scriptscriptstyle\bullet}="B";
     (3,-1)*{\scriptscriptstyle\bullet}="C";
     "A";"B"**\dir{-};
     "C";"B"**\dir{-};
     "A";"C"**\dir{-};
\endxy
\) 
are labeled by `irreducible intertwiners', for example going from the tensor product 
\(
\xy
     (-3,-1)*{\scriptscriptstyle\bullet}="A";
     (0,1)*{\scriptscriptstyle\bullet}="B";
     (3,-1)*{\scriptscriptstyle\bullet}="C";
     "A";"B"**\dir{-};
     "C";"B"**\dir{-};
     (-2,1.5)*{\scriptscriptstyle X\phantom{'}};
     (3,1.5)*{\scriptscriptstyle X'};
\endxy
\)
to the single irreducible representation 
\(
\xy
     (-3,-1)*{\scriptscriptstyle\bullet}="A";
     (0,1)*{\phantom{\scriptscriptstyle\bullet}}="B";
     (3,-1)*{\scriptscriptstyle\bullet}="C";
     "A";"C"**\dir{-};
     (0,.5)*{\scriptscriptstyle Y};
\endxy
\). 
To describe such intertwiners, consider the set 
\[
T 
  =\{ (x,x',y)\in X\times X'\times Y : x +x' = y\}
\]
of ways to build a triangle using one vector from each orbit $X,X',Y$. The diagonal action of $\SO(4)$ on $X\times X' \times Y$ restricts to an action on $T$, which we write $(g, \vartriangle) \mapsto g \!\!\vartriangle$.  
An intertwiner amounts to an {\bf\boldmath $\SO(4)$ Hilbert bundle} over $T$, that is, a vector bundle $V$ whose fibers are Hilbert spaces, with a fiber-preserving $\SO(4)$ action. More precisely, for $\vartriangle\in T$, $\varphi \in V_{\!\vartriangle}$, and $g \in \SO(4)$, we can write the action as 
$g (\!\vartriangle, \varphi)  = (g\!\!\vartriangle, \Phi^g_{\!\vartriangle}(\varphi))$, where $\Phi^g_{\!\vartriangle}\maps V_{\!\vartriangle}\to V_{\! g \!\vartriangle}$ are linear maps satisfying
\begin{equation}
\label{cocycle}
\Phi^{gg'}_{\!\vartriangle} = \Phi^g_{g \!\vartriangle} \Phi^{g'}_{\!\vartriangle}
\end{equation}
This equation says {\em sections} of the vector bundle $V$ form a representation of $\SO(4)$.   Using an $\SO(4)$-invariant measure $\mu$ on $T$, we can restrict to {\bf\boldmath $L^2$ sections} $f$, for which $\int_T \extd \mu(\vartriangle) |f(\vartriangle)|^2 < \infty$, and get a {\em unitary} representation of $\SO(4)$. Alternatively, if we {\em fix} a reference point $\!\vartriangle \in T$ and restrict the action to its stabilizer $G_{\!\vartriangle} \subseteq \SO(4)$, so $g\!\!\!\vartriangle = \!\vartriangle$, then (\ref{cocycle}) is just the equation for a representation of $G_{\!\vartriangle}$ on $V_{\!\vartriangle}$.   Moreover, the full $\SO(4)$ representation, hence the 2-group intertwiner, can be reconstructed, up to equivalence, from this $G_{\!\vartriangle}$ representation---it is just an `induced representation' \cite{Mackey1978}. The intertwiner turns out to be {\bf irreducible} if and only if the $G_{\!\vartriangle}$ representation is irreducible. 
If the radii labeling the three edges satisfy the triangle inequality, $G_{\!\vartriangle}$ is isomorphic to $\U(1)$, so we may think of triangles as labeled by elements of $\Irrep(\U(1))\iso \Z$.

{\bf Gluing triangles.}  Triangles with a common edge label, like
\(
\xy
     (-3,-1)*{\scriptscriptstyle\bullet}="A";
     (0,1)*{\scriptscriptstyle\bullet}="B";
     (3,-1)*{\scriptscriptstyle\bullet}="C";
     "A";"B"**\dir{-};
     "C";"B"**\dir{-};
     "A";"C"**\dir{-};
     (-2,1.5)*{\scriptscriptstyle X\phantom{'}};
     (3,1.5)*{\scriptscriptstyle X'};
     (0,-1)*{\color{white}\rule{.5em}{.5em}};
     (0,-1)*{\scriptscriptstyle Y};     
\endxy
\) 
and 
\(
\xy
     (-3,1.5)*{\scriptscriptstyle\bullet}="A";
     (0,-.5)*{\scriptscriptstyle\bullet}="B";
     (3,1.5)*{\scriptscriptstyle\bullet}="C";
     "A";"B"**\dir{-};
     "C";"B"**\dir{-};
     "A";"C"**\dir{-};
     (-2,-.5)*{\scriptscriptstyle Z\phantom{'}};
     (3,-.5)*{\scriptscriptstyle Z'};
     (-.1,1.5)*{\color{white}\rule{.5em}{.2em}};
     (0,2)*{\scriptscriptstyle Y};     
\endxy
\),
 can be glued together: 
\(
\xy 
     (-3,0)*{\scriptscriptstyle\bullet}="A";
     (0,1.8)*{\scriptscriptstyle\bullet}="B";
     (3,0)*{\scriptscriptstyle\bullet}="C";
     (0,-1.8)*{\scriptscriptstyle\bullet}="D";
     "A";"B"**\dir{-};
     "C";"B"**\dir{-};
     "A";"D"**\dir{-};
     "C";"D"**\dir{-};
     "A";"C"**\dir{-};
\endxy
\)
to give more complicated, generally `reducible', intertwiners.  
Geometrically, the resulting 2-intertwiner from $X\tensor X'$ to $Z\tensor Z'$ is an $\SO(4)$ Hilbert bundle over the set
\[
  Q 
  =  \{(x,x',z,z')\in X\times X' \times Z \times Z' : x+x' = z+z' \in Y \}.
\]
of ways of constructing the shape
\(
\xy 
     (0,.5)="shift";
     "shift"+(-3,0)*{\scriptscriptstyle\bullet}="A";
     "shift"+(0,1.8)*{\scriptscriptstyle\bullet}="B";
     "shift"+(3,0)*{\scriptscriptstyle\bullet}="C";
     "shift"+(0,-1.8)*{\scriptscriptstyle\bullet}="D";
     "A";"B"**\dir{-};
     "C";"B"**\dir{-};
     "A";"D"**\dir{-};
     "C";"D"**\dir{-};
     "A";"C"**\dir{-};
\endxy
\)
out of vectors in $\R^4$ with lengths fixed by the edge labels. Note that this shape is not `rigid' in $\R^4$, since it can be bent along the joining edge without changing any edge lengths, so there are many orbits in this set. In terms of the intertwiners $(V_{\utriangle}, \Phi^g_{\utriangle})$ and $(V_{\,\!\dtriangle}, \Phi^g_{\!\dtriangle})$ labeling the two triangles, this bundle assigns the vector space $W_{\hdiamond} := V_{\utriangle} \otimes V_{\,\!\dtriangle}$ to each element $\hdiamond \in Q$. An element $g \in \SO(4)$ acts on this bundle as $g (\hdiamond, \varphi) = (g \hdiamond, \Psi^g_{\hdiamond}(\varphi)) $, with $\Psi^g_{\hdiamond} := \Phi^g_{\utriangle} \otimes \Phi^g_{\!\dtriangle}$. For generic edge labels satisfying the triangle inequality, all vector spaces are 1-dimensional, so we really have a complex $\SO(4)$ {\em line} bundle.

{\bf Tetrahedra: 2-intertwiners.}  A tetrahedron 
\(\xy 
     (-3,0)*{\scriptscriptstyle\bullet}="A";
     (0,1.8)*{\scriptscriptstyle\bullet}="B";
     (3,0)*{\scriptscriptstyle\bullet}="C";
     (0,-1.8)*{\scriptscriptstyle\bullet}="D";
     "A";"B"**\dir{-};
     "C";"B"**\dir{-};
     "A";"D"**\dir{-};
     "C";"D"**\dir{-};
     {\knotholesize{4pt}}
     \xoverv~{"A"}{"B"}{"D"}{"C"};
\endxy
\) is labeled by a `2-intertwiner' relating its back: 
\(
\xy 
     (-3,0)*{\scriptscriptstyle\bullet}="A";
     (0,1.8)*{\scriptscriptstyle\bullet}="B";
     (3,0)*{\scriptscriptstyle\bullet}="C";
     (0,-1.8)*{\scriptscriptstyle\bullet}="D";
     "A";"B"**\dir{-};
     "C";"B"**\dir{-};
     "A";"D"**\dir{-};
     "C";"D"**\dir{-};
     "A";"C"**\dir{-};
\endxy
\) to its front:
\(
\xy 
     (-3,0)*{\scriptscriptstyle\bullet}="A";
     (0,1.8)*{\scriptscriptstyle\bullet}="B";
     (3,0)*{\scriptscriptstyle\bullet}="C";
     (0,-1.8)*{\scriptscriptstyle\bullet}="D";
     "A";"B"**\dir{-};
     "C";"B"**\dir{-};
     "A";"D"**\dir{-};
     "C";"D"**\dir{-};
     "B";"D"**\dir{-};
\endxy
\), both constructed as in the previous paragraph.  
If the horizontal edge is labeled by the representation $Y$, and the vertical one by $Y'$, we can view both front and back as intertwiners from $X\tensor X'$ to $Z\tensor Z'$, but subject to different triangle gluing constraints:
\[
   x + x' = z + z' \in Y \quad {\rm and} \quad x+z=x'+z' \in Y'
\]
respectively. A 2-intertwiner for the tetrahedron is a {\bf\boldmath map of $\SO(4)$ vector  bundles}, {\em restricted} to the intersection of their domains, namely the space of `tetrahedra':
\[
  \cT =  \{(x,x',z,z')\in X\times X' \times Z \times Z' : x+x' = z+z' \in Y \; {\rm and} \; 
           x + z = x' + z' \in Y' \}
\]
Any two tetrahedra in $\R^4$ with the same edge lengths differ by an orientation-preserving isometry, so there is just one $\SO(4)$ orbit in this space of tetrahedra.  
In terms of the intertwiners $(V_{\utriangle}, \Phi^g_{\utriangle})$, $(V_{\,\!\dtriangle}, \Phi^g_{\!\dtriangle})$, $(V_{\ltriangle}, \Phi^g_{\ltriangle})$ and $(V_{\rtriangle}, \Phi^g_{\rtriangle})$ labeling the four triangles of the tetrahedron, the map of vector bundles amounts to a family of maps  $m^{}_{\hvdiamond}$ from $V_{\utriangle} \otimes V_{\!\dtriangle}$ to 
$V_{\ltriangle} \otimes V_{\rtriangle}$, labeled by elements $\hvdiamond \in \cT$, and satisfying the intertwining property:
\[
\Phi^g_{\utriangle} \otimes  \Phi^g_{\!\dtriangle} \, m^{}_{\hvdiamond} = 
 m^{}_{g\hvdiamond} \, \Phi^g_{\ltriangle} \otimes \Phi^g_{\rtriangle}
\]
As with the equation for triangles (\ref{cocycle}), this can be interpreted in two ways: either as an ordinary intertwiner between $\SO(4)$ representations on sections, or pointwise as intertwiners of stabilizer representations. 

The representation theory for an arbitrary (strict skeletal) 2-group $\G=(G,H)$ is similar in spirit to that just described for $\E$.  We have, however, taken advantage of some nice features of $\E$, and it seems appropriate here to indicate briefly what may differ in the general case. First, the Pontrjagin dual $\Hhat = \hom(H,U(1))$ of $H$ plays a crucial role; this is easily overlooked for $\G=\E$ since $\R^4$ is isomorphic to its dual.  Irreducible representations are in general $G$-orbits in $\Hhat$.  
Also, the action of $G$ on the analog of the set of `triangles' $T$ need not be transitive in general. An irreducible intertwiner is then a bundle over a single orbit in $T$; 
thus triangles carry in general additional labels corresponding to $G$-orbits in $T$.  We have also glossed over some measure-thoretic issues that can be finessed away for $\E$, but are vital in other important examples, including the Poincar\'e 2-group.

We can summarize the essential ingredients in the general case as follows:
{\small
\begin{center}
\begin{tabular}{c|c|c}
 &
     label   &   geometric characterization
    \\     \hline 
\(
\xy
     (-3,-1)*{\scriptscriptstyle\bullet}="A";
     (3,-1)*{\scriptscriptstyle\bullet}="C";
     "A";"C"**\dir{-};
     (0,.3)*{\scriptscriptstyle X};
\endxy
\)  
&    
\txt{irreducible \\ representation}  
&  
{\bf\boldmath $G$-orbit in $\Hhat$}
\\
\hline
\(
\xy
     (-3,-1.5)*{\scriptscriptstyle\bullet}="A";
     (0,1.5)*{\scriptscriptstyle\bullet}="B";
     (3,-1.5)*{\scriptscriptstyle\bullet}="C";
     "A";"B"**\dir{-};
     "C";"B"**\dir{-};
     "A";"C"**\dir{-};
     (-2,1.5)*{\scriptscriptstyle X\phantom{'}};
     (3,1.5)*{\scriptscriptstyle X'};
     (0,-3)*{\scriptscriptstyle Y};     
\endxy
\)  
&  \txt{irreducible \\ intertwiner \\ from $X\tensor X'$ to $Y$}   
&  \txt{$G$-orbit $\cO$ in the`space of triangles'  in $\Hhat$,  \\  with a {\bf\boldmath $G$ Hilbert bundle} over $\cO$ \\ fiber over $\!\vartriangle$ = irreducible representation of  stabilizer $G_{\!\vartriangle}$} 
\\[1.3em]
\hline
\(\xy 
     (-5,0)*{\scriptscriptstyle\bullet}="A";
     (0,5.5)*{\scriptscriptstyle\bullet}="B";
     (5,0)*{\scriptscriptstyle\bullet}="C";
     (0,-3)*{\scriptscriptstyle\bullet}="D";
     "A";"B"**\dir{-};
     "C";"B"**\dir{-};
     "A";"D"**\dir{-};
     "C";"D"**\dir{-};
     {\knotholesize{4pt}}
     \xoverv~{"A"}{"B"}{"D"}{"C"};
     (-3,3.5)*{\scriptscriptstyle X\phantom{'}};
     (4,3.5)*{\scriptscriptstyle X'};
     (-2.2,.2)*{\color{white}\rule{.3em}{.5em}};
     (-2,.2)*{\scriptscriptstyle Y};          
     (.5,2)*{\color{white}\rule{.5em}{.7em}};
     (.5,2.2)*{\scriptscriptstyle Y'};          
     (-3,-2)*{\scriptscriptstyle Z\phantom{'}};
     (4,-2)*{\scriptscriptstyle Z'};
\endxy
\)
& \txt{2-intertwiner \\ \rule{0em}{1.2em}
\(
\xy 
     (-3,0)*{\scriptscriptstyle\bullet}="A";
     (0,1.8)*{\scriptscriptstyle\bullet}="B";
     (3,0)*{\scriptscriptstyle\bullet}="C";
     (0,-1.8)*{\scriptscriptstyle\bullet}="D";
     "A";"B"**\dir{-};
     "C";"B"**\dir{-};
     "A";"D"**\dir{-};
     "C";"D"**\dir{-};
     "A";"C"**\dir{-};
\endxy
\To
\xy 
     (-3,0)*{\scriptscriptstyle\bullet}="A";
     (0,1.8)*{\scriptscriptstyle\bullet}="B";
     (3,0)*{\scriptscriptstyle\bullet}="C";
     (0,-1.8)*{\scriptscriptstyle\bullet}="D";
     "A";"B"**\dir{-};
     "C";"B"**\dir{-};
     "A";"D"**\dir{-};
     "C";"D"**\dir{-};
     "B";"D"**\dir{-};
\endxy
\)}    
&  {\bf\boldmath map of $G$ Hilbert bundles} 
\end{tabular}
\end{center}
} 
\noindent Details about the general representation theory can be found in \cite{2Rep}.

\section{The Euclidean 2-group model}

Knowing the representation theory, and how to use it to label cells of a triangulation $\Delta$ of some manifold, the key remaining step in writing a spin foam model for the Euclidean 2-group $\E$ is assigning appropriate weights $W_\Delta(s_t,\ell_e)\in \C$ depending on the length and spin labels $\ell_e\in \R^+, s_t\in \Z$ of $\Delta$. 
An explicit model is developed in \cite{BaratinFreidel2}, where these weights are
\[
W_{\Delta} (s_t, \ell_e)  = \prod_{t \in \Delta} {{\rm A}_t(\ell_e)} \prod_{\sigma \in \Delta}\, \frac{\cos S_\sigma(s_t, \ell_e)}{\rm V_\sigma(\ell_e)} 
\]
Here ${\rm A}_t(\ell_e)$ is the area of the triangle $t$ computed from the edge lengths $\ell_e$. Each 4-simplex $\sigma$ gets a factor involving its volume $V_\sigma(\ell_e)$ 
and the `first order Regge action' $S_\sigma = \sum_{t\in \sigma} s_t \omega_t^\sigma$, where $\omega_t^\sigma$ is the dihedral angle of $t$ in the 4-simplex $\sigma$.

Let us briefly describe how the weights $W_\Delta$ are obtained. 
First, we have seen that a face label can be viewed as a unitary representation of $\SO(4)$---the $L^2$ sections of a line bundle over a space $T$ of triangles in $\R^4$ with fixed lengths.  Defining $L^2$ sections required choosing a suitable measure $\mu$ on $T$ for each triangular face $t$.  Any invariant measure will work, but it must be properly normalized.  With a natural choice of measures coming from the geometry, this normalization is just the {\em area} of $t$, up to an overall factor, so this gives us the `{\em face} factors' in the $W_\Delta$.  Similarly, we have seen that a 2-intertwiner is a map of $\SO(4)$ line bundles over a space $\cT$ of tetrahedra with fixed edge lengths. Given a suitable measure on $\cT$, it gives an ordinary intertwiner relating the unitary representations of $\SO(4)$ on the boundary faces. The `{\em 4-simplex} factor' is defined by taking the trace of the product of five $\SO(4)$ intertwiners on the bounding tetrahedra. The result is a `20j-symbol'---a function of ten edge labels and ten face labels \cite{BaratinFreidel2}.

Given these weights, the model is:
\beq \label{statesum}
Z_{\Delta} = \int \, \prod_{e \in \Delta} \ell_e d\ell_e \sum_{s_t \in \Z} \, W_{\Delta} (s_t, \ell_e) 
\eeq
From the perspective of 2-group representation theory, it is very natural to propose models of this sort. Our interest here in such models, however, is not purely mathematical: the Euclidean 2-group model shows up in a surprising but natural way in physics.  We now turn to explaining this.

\section{From Feynman graphs to `quantum flat space'}

The model just described was first obtained without presupposing any 2-group structure \cite{BaratinFreidel1b}.  The goal of this work was simply to rewrite standard Feynman amplitudes
\[
\int d^4x_1 \ldots d^4x_n \prod_{(ij)\in \Gamma} G(x_i - x_j)
\]
of Euclidean scalar field theory in the combinatorial language of state sums. It was shown that the {\em measure} used to evaluate the integral in Feynman amplitudes {\em is} in fact a state sum model, precisely the one given by (\ref{statesum}).  

The strategy for showing this is to gauge out the Poincar\'e symmetry of the integrand, favoring the distances 
$\ell_{ij} := \| x_i - x_j \|$ as variables, over the vertex positions $x_i$. One then views the $\ell_{ij}$ as providing {\em flat} Regge geometries of the 4-sphere $\cS^4$, for a certain class of triangulations built from the Feynman graph $\Gamma$. With more work, the construction can be extended to sum over {\em all} Regge geometries of $\cS^4$, with flatness imposed by delta functions forcing the deficit angle $\omega$ at each triangle to vanish. Fourier expanding these delta functions \(\delta(\omega) \sim  \sum_{s\in \Z} \exp(i s \omega)\) gives additional spin variables $s_t \in \Z$ labeling triangular faces $t$. 

The end result is that Feynman amplitudes can be computed from the observables
\[
I_{\Gamma} = \int \, \prod_{e \in \Delta} \ell_e d\ell_e \sum_{\{s_t\}} \, W_{\Delta} (s_t, \ell_e)  \prod_{e \in \Gamma} G(\ell_e)
\]
obtained by coupling the product of Feynman propagators $\prod_{e \in \Gamma} G(\ell_e)$  to the fluctuating geometries $\{\ell_e\}$ of the model (\ref{statesum}). 
Here $\Delta$ is any triangulation of $\cS^4$ having the graph $\Gamma$ as a subcomplex. Certain identities satisfied by the state sum weight insure that the observables do not depend (after suitable gauge-fixing) on the chosen triangulation $\Delta$. 
 
This result gives a background independent perspective on quantum field theory amplitudes. The state sum can be viewed as a `quantum model of flat space', where flatness is implemented dynamically by the choice of the quantum weight.

\section{Beyond the flat model}

In the main model we have described, the Euclidean {\em group} $\SO(4)\ltimes \R^4$ is reinterpreted as a {\em 2-group}, leading to a new sort of representation theory to be used in the state sum. Of course, this is really meant to be a warmup to an analogous Lorentzian model based on the Poincar\'e 2-group.

It is tempting to relate this model to a Poincar\'e gauge theory for gravity. However, the most convincing such gauge theory---the MacDowell Mansouri formulation \cite{MM,Wise,Wise2}---has not the Poincar\'e group but the {\em de Sitter group} $\SO(4,1)$ as gauge group.  While the de Sitter group cannot be viewed as a 2-group like the Poincar\'e group can, it is possible to instead `cosmologically deform' the {\em representation 2-category} of the Poincar\'e 2-group.  Guided by the geometric description in the Euclidean or Poincar\'e case, in such a `de Sitter deformation', for example, irreducible representations become $\SO(3,1)$ orbits, not in Minkowski space, but in de Sitter, while 1- and 2-intertwiners involve gluing relations for triangles and tetrahedra living in de Sitter space.  Analogously with the Poincar\'e case, the resulting state sum involves `Regge geometries' whose simplices are only `flat' in Cartan's generalized sense of being isometric with a portion of de Sitter space \cite{BahrDittrich,Wise,Wise2}. The positive-signature analog of such a model shows up in a state sum formulation of Feynman amplitudes for quantum field theory on spherical space \cite{BaratinFreidel1b}.

But other generalizations of the model presented here may also be interesting.  In principle, one should be able to develop analogous models not only for other {\em strict skeletal} 2-groups $(G,H)$, but also more general 2-groups, whose representation theory is still not fully understood.  These may be interesting from the point of view of quantum topology.  While the Euclidean 2-group model is formally triangulation independent \cite{BaratinFreidel1b}, we do not yet know whether this property is special, or common to a wide class of 2-group state sum models.  If triangulation independence {\em is} generic,  interesting invariants might be obtained by making a good choice of 2-group. 

On a more physical side, a better understanding of the geometric content of these models may give some guiding insights for realistic models of quantum geometry.  We emphasize that the perspective on Feynman amplitudes presented in the previous section was originally motivated by background independent approaches to quantum gravity. Recent results in three dimensional spin foam gravity \cite{FreidelLivine} have led to the heuristic idea that quantum gravity should provide a {\em measure} for the integrals in Feynman graph amplitudes ({\it cf.} \cite{Barrett}). When gravity is `turned off', this measure should be the standard Lebesgue measure. With gravity `on', the Lebesgue measure is deformed and should take into account quantum geometry corrections.  On the other hand, the spin foam approach says quantum gravity should be described by a background-free state sum model. This motivated the attempt \cite{BaratinFreidel1a, BaratinFreidel1b} to reformulate Feynman amplitudes as background-free state sums. 

Taking this idea seriously suggests that the state sum structure shown here, hence the Poincar\'e 2-group model, may contain some seed of information about the structure of the quantum gravity amplitude itself.


\begin{theacknowledgments}
We thank John Baez and Laurent Freidel for collaboration on \cite{2Rep,BaratinFreidel1b,BaratinFreidel2}, as well as Jeffrey Morton for helpful discussions.  D.W.\ was supported in part by the National Science Foundation under grant DMS-0636297.   
\end{theacknowledgments}

\vskip 3.8cm
\begin{flushright}  AEI-2009-100 \end{flushright}

\end{document}